# COULOMB WAVE DISCRETE VARIABLE METHOD FOR CALCULATION OF THE FULLY DIFFERENTIAL CROSS SECTIONS OF ANTIPROTON IMPACT IONIZATION OF HYDROGEN ATOM

G. Zorigt*, L. Khenmedekh, Ch. Aldarmaa

*Department of Physics, School of Applied Sciences, Mongolian Science and Technology University*

*E-mail: g_zorigt@yahoo.com*

**Abstract**

We studied nonrelativistic collision of antiproton with hydrogen atom by solving time-dependent Schrodinger equation numerically. Coulomb wave function discrete variable method (CWDVR) had been used to calculate electron wave function evolution, while projectile defined classically, moving along the straight line trajectories with constant velocity. The ionization amplitude calculated by projection of the wave function into continuum wave function of the hydrogen electron. The fully differential cross sections calculated depending on projectile impact energy, scattering angle and electron ejection energy and angles. Our results in good agreement with the relativistic calculation results.

## 1. Introduction

Theory of charged particle collision had been studied from establishment of quantum theory in the middle of 1920s, beginning from the Born theory of collision. Experimental equipment's are continuously improved, and can accurately measure the differential cross sections of the collision. From the computational point of view, the antiproton collision with Hydrogen atom is simplest case, because against the proton case, the electron capture channel doesn't contribute to the collision cross section.

In the resent years, few theoretical works among them perturbative theory of Voitkiv et al [1] based on Continuum Distorted Wave (CDW) approximation. At low energy of projectile, for few hundred keV, the perturbation theory become inaccurate, and non perturbative theory can be applied. Recently, developed few non preturbative theories, treating the projectile classically, as moving along a straight line, but the atomic electrons treated quantum mechanically. Among them, coupler pseudo state (CP) theory of McGovern et al [2,3], time dependent convergent – close – coupling (QM - CCC) approach (Abdurakhmanov et al [4]) , time dependent close coupling (TDCC) method (Ciappina et al [5]) could be mentioned. Recently, Boudarev et al [6] used the relativistic single center semi classical coupled channel approach had been developed to calculate fully differential cross section (FDCS).

In the present study, for the first time, the Coulomb wave functions have been used as a basis in the discrete variable representation (L.Y. Peng, A.F Starace [7]) to study the antiproton-hydrogen collisions. These wave functions have been first expanded into the spherical harmonic functions as they enable to explicitly calculate the time-dependent radial functions. By considering the classical assumption that an antiproton with a particular impact parameter preserves its straight trajectory, we show how the hydrogen electron has temporally evolved from its ground state in the field produced by the presence of the antiproton and compared the results of the FDCS to the previously published theoretical data. In particular, we compare our results for the FDCS-s in the antiproton-atom collision, in which an initially 2 – 10 MeV energetic antiproton is scattering angle 10-40mkrad and electron ejection energy 5eV. to the previously published results.

## 2. Theory

Assuming the antiproton moving along the straight trajectory, the atomic electron excitation and/or ionization processes are studied here as functions of the antiproton impact parameter. We adopt numerical approach constructing a space net and calculate the wave function on each knot of the net in the spherical coordinate system. This is accomplished by solving the one of the fundamental equations of the quantum mechanics – the time-dependent Schrödinger equation. Namely, the Schrödinger equation for an electron of the hydrogen atom in the external field is given by

$$i\frac{\partial \Psi(\vec{r},t)}{\partial t} = \left(\hat{H}_0 + \hat{V}(\vec{r},t)\right)\Psi(\vec{r},t) \qquad (1)$$

where $\Psi(\vec{r},t)$ is wave function and $\hat{H}_0$ is the Hamiltonian for an electron of a hydrogen atom, $\hat{V}(\vec{r},t)$ is the interaction opreator due to the external field. Assuming that the antiproton or ion moves along the straight line (z-axis), the produced electric potential is expressed as follows

$$\hat{V}(r,t) = -\frac{1}{|R(b,0,vt)-r|} \quad (2)$$

here $R$ is ion radius vector, $b$ is impact parameter, $v$ is ion velocity, $r$ is electron radius vector, t is time. With this assumption that the antiproton moves along the classical trajectory, the ionization and scattering calculations have been performed in the cases of antiproton-helium [8-10] and antiproton-hydrogen [6] collisions.

The fully differential cross section is the one of the experimentally verifiable quantities as results of ionization processes. The ionization amplitude is obtained by an overlap integral constructed from the wave functions calculated in the discrete variable representation as

$$T(\varepsilon, \theta_e, \varphi_e, b, \varphi_b) = \langle \psi^- | \psi_f \rangle \quad (3)$$

where for impact parameter $\vec{b}(b, \varphi_b)$, an electron ionization amplitude for scattering angle $\theta_e, \varphi_e$ and energy $\varepsilon$ is $T(\varepsilon, \theta_e, \varphi_e, b, \varphi_b)$ and $\psi^-$ is continuum wave function, $\psi_f$ is wave function at a final moment in the calculation.

Next, the ionization amplitude is converted from its impact parameter representation to the momentum transport representation by two-dimensional Fourier transform as in Bondarev et al. [3].

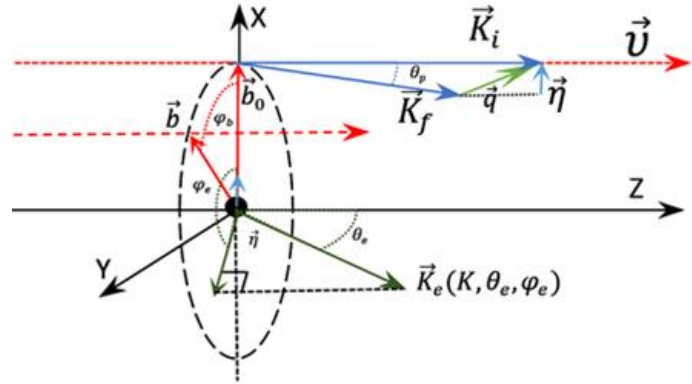

Figure 1. *Antiproton enters along z-axis. $\vec{K}_i, \vec{K}_f$ are initial and final momenta of the antiproton $\vec{K}_e$ is the electron momentum, $\vec{\eta}$ is the component perpendicular to the antiproton velocity $\vec{v}$ of the momentum transport from the antiproton to the atom.*

The geometric construction of the vectors for impact parameter and momentum transport is illustrated schematically in Fig. 1. As shown in figure 1, the ionization amplitude is obtained by the integral given in Eq. (4) in the representation of $\vec{\eta}$ - the perpendicular component of the momentum transfer to the antiproton's velocity $\vec{v}$ as follows,

$$T(\varepsilon, \theta_e, \varphi_e, \eta, \varphi_\eta) = \frac{1}{2\pi} \int d\vec{b} e^{i\vec{\eta}\vec{b}} e^{i\delta(b)} T(\varepsilon, \theta_e, \varphi_e, b, \varphi_b) \quad (4)$$

here $\delta(b)$ is the phase shift accounting the ion-atom interaction as given by

$$\delta(b) = \frac{2Z_P Z_T}{v} \ln(vb) \quad (5)$$

Due to the momentum transfer $\vec{\eta}(\eta, \varphi_\eta)$ from the antiproton, assume that the electron is ionized with an energy of $\varepsilon$ with angle $\theta_e, \varphi_e$. The corresponding ionization fully differential cross section is expressed as

$$\frac{d^3\sigma}{d\varepsilon d\Omega_e d\Omega_P} = K_i K_f |T(\varepsilon, \theta_e, \varphi_e, \eta, \varphi_\eta)|^2 \quad (6)$$

where $d\Omega_p$ is differentiation of the scattering solid angle for the antiproton, $K_i, K_f$ are initial and final momenta of the antiproton. In the relative coordinate system, the atomic nucleus is assumed to be at rest and antiproton is expressed with its reduced mass.

## 3. Results and Discussion

We consider the antiproton collision with Hydrogen atom in it is ground state. First, we calculated the ionization amplitude for projectile flying along straight-line trajectory with constant velocity and for given impact parameter. Impact parameter b varies with in the integral (0.1-100 a.u.). We used symmetry properties (axial symmetry around the Z- axis, which is parallel to projectile trajectory and mirror symmetry with respect to the scattering plane.) to economy the calculations. The ionization amplitudes for different impact parameters are collected as a data for further calculations of scattering amplitude with given projectile scattering angle using formula (4), in which the integrals calculated numerically.

Then fully differential cross sections are obtained via the formula (6). We perform calculations for different projectile energies and different scattering angles. Figure 2a shows differential cross sections of ionization for projectile with energy of 200 keV and scattering angle of 0.2 mrad, where electron ejection energy is equal to 4 eV. Next figure 2b shows differential cross sections of ionization for projectile with energy of 500 keV and scattering angle of 0.024 mrad, where electron ejection energy is equal to 5 eV.

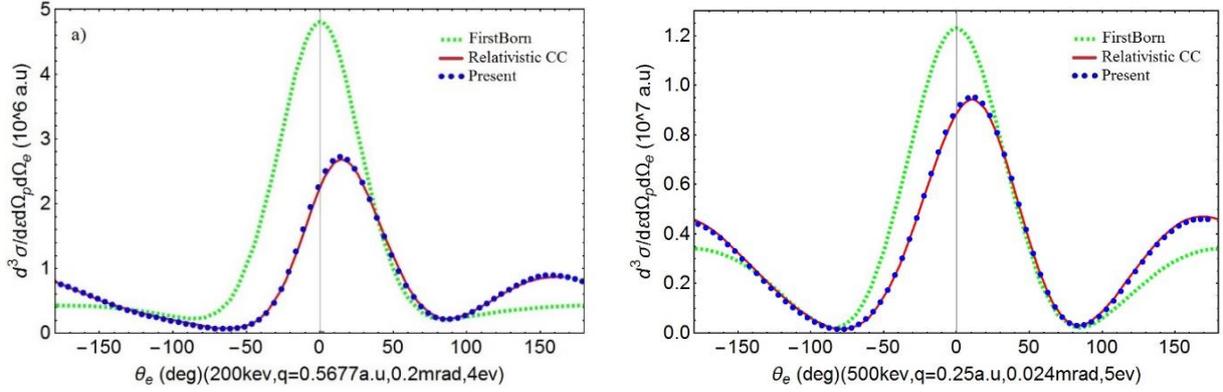

Figure 2. *FDCS in the scattering plane. The projectile energy is equal to a). 200 keV and b).500keV . The results of the relativistic coupled-channel are from Refs.[4,6]*

As it seen from the Fig. 2, our calculations are in agreement with the relativistic cross sections of A.I.Bondarev et al [6] and CCC results of I.B. Abdurakhmanov et.al. [4]. Our results in good agreement with the relativistic calculation results. At this impact energies, the first Born cross section binary peak is much higher in comparison with our calculated result, but the recoil peak is much smaller.

For high incident energy of the projecttile, first Born approximation(FBA) will be applicable. So we calculated FDCS for the energy values 2, 5 and 10MeV, and compared with corresponding FBA results.

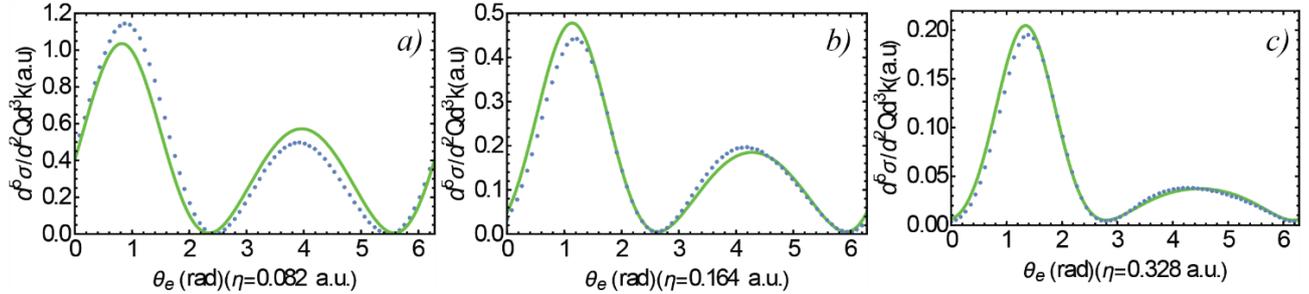

Figure 3. *FDCS in the scattering plane. The collision parameters:The projectile energy of 2MeV and scattering angle is equal to a). 10 µrad  b). 20 µrad and c). 40 µrad, where electron ejection energy is equal to 5eV and η - the momentum transfer.  Line for first Born approximation, dot for result CWDVR.*

For the 2MeV incident energy,CWDVR and  FBA cross sections are have nearly identical shapes, but as seen from fig.3. the hight and angular positions of the binary and recoil peaks are show a littele difference.

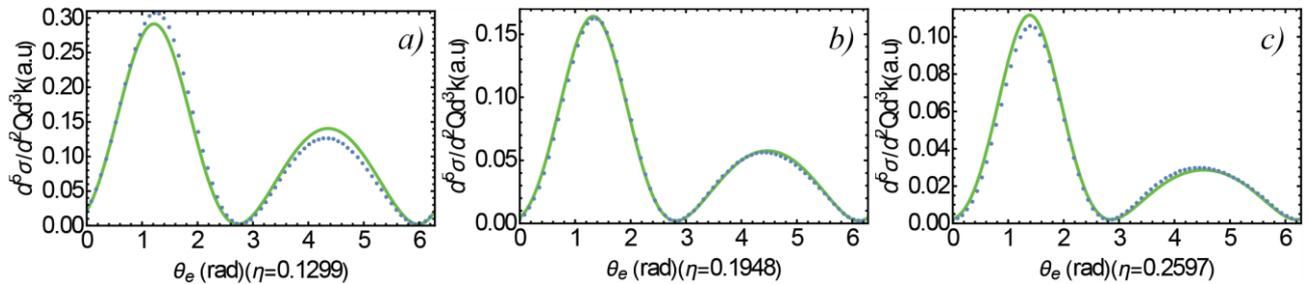

Figure 4. *FDCS in the scattering plane. The collision parameters : The projectile energy of 5MeV and scattering angle is equal to a). 10 µrad b). 15 µrad and c). 20 µrad , where electron ejection energy is equal to 5eV and η - the momentum transfer.  Line for first Born approximation, dot for result CWDVR.*

For the 5MeV incident energy, differences in peak hight and angular positionsare remain,but as seen from fig.3. for about 15μrad scattering angle the two cross sections are coincide.

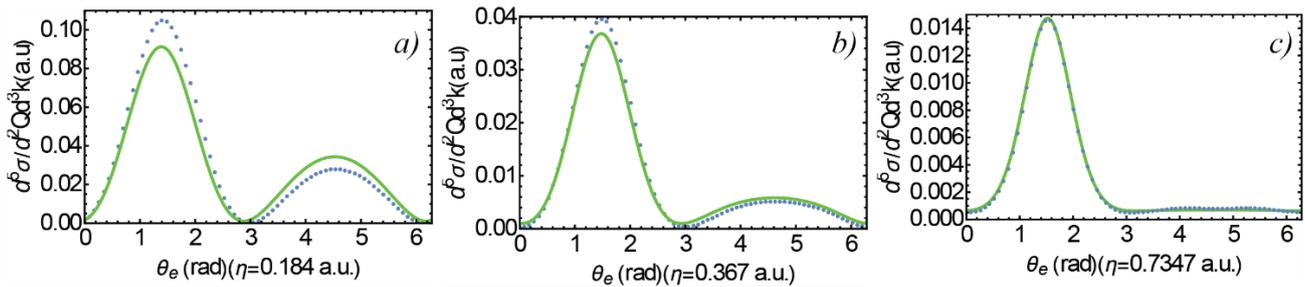

Figure 5. *FDCS in the scattering plane.The collision parameters:The projectile energy of 10MeV and scattering angle is equal to a). 10 μrad b). 20 μrad and c). 40 μrad., where electron ejection energy is equal to 5eV and η - the momentum transfer. Line for first Born approximation, dot for result CWDVR.*

The angular positions for FBA and CWDVR cross sections are coincide for the 10 MeV incident energy and nearly perfect coincidence of FDCS observed for the 40 *μrad scattering angle.*

## 4. Conclusions

Fully differential ionization cross sections of antiproton collisions with Hydrogen atom are obtained from the numerical solutions of the time dependent Schrödinger equation using CWDVR method.
We test our numerical method for different incident energiys and scattering angles of the projectile and get convergence to the relativistic cross sections of A.I.Bondarev et al [6] and CCC results of I.B. Abdurakhmanov et.al. [4].For higher incident energies of 2, 5, and 10 MeV CWDVR FDCS results are close to the FBA results and coinsides with them for special scattering angles.